\documentstyle[12pt]{article}
\begin{document}
\large
\bibliographystyle{plain}

\begin{titlepage}
\hfill \begin{tabular}{l} HEPHY-PUB 606/94\\ UWThPh-1994-23\\ June 1994
\end{tabular}\\[4cm]
\begin{center}
{\Large\bf A VARIATIONAL APPROACH TO THE SPINLESS RELATIVISTIC COULOMB
PROBLEM}\\
\vspace{2.5cm}
{\Large\bf Wolfgang LUCHA}\\[.5cm]
{\large Institut f\"ur Hochenergiephysik\\
\"Osterreichische Akademie der Wissenschaften\\
Nikolsdorfergasse 18, A-1050 Wien, Austria}\\[1cm]
{\Large\bf Franz F. SCH\"OBERL}\\[.5cm]
{\large Institut f\"ur Theoretische Physik\\
Universit\"at Wien\\
Boltzmanngasse 5, A-1090 Wien, Austria}\\[2.5cm]
{\bf Abstract}
\end{center}
\normalsize

\noindent
By application of a straightforward variational procedure we derive a simple,
analytic upper bound on the ground-state energy eigenvalue of a
semirelativistic Hamiltonian for (one or two) spinless particles which
experience some Coulomb-type interaction.
\end{titlepage}

\section{Introduction}

A standard (semi-) relativistic description of bound states built up by
particles of spin zero is provided by a Hamiltonian $H$ which
\begin{itemize}
\item incorporates relativistic kinematics by involving the
{}\kern-.10em\hbox{``}square-root" operator of the relativistic kinetic
energy, $\sqrt{\vec p\,{}^2 + m^2}$ for a particle of mass $m$ and momentum
$\vec p$, but
\item describes the forces acting between the bound-state constituents by
some coordinate-dependent static interaction potential $V(\vec x)$.
\end{itemize}
For the case of bound states consisting of two particles of equal mass $m$,
the generic Hamiltonian in the center-of-momentum frame of these
constituents, expressed in terms of relative momentum $\vec p$ and relative
coordinate $\vec x$, reads
\begin{equation}
H = 2\sqrt{\vec p\,{}^2 + m^2} + V(\vec x) \quad .
\label{eq:semrelham}
\end{equation}
The equation of motion resulting from this type of Hamiltonian is the
well-known {}\kern-.10em\hbox{``}spinless Salpeter equation." It may be
obtained from the so-called Salpeter equation \cite{salpeter52} by ignoring
all spin degrees of freedom and considering positive-energy solutions only.
The Salpeter equation, in turn, approximates the Bethe--Salpeter equation
\cite{salpeter51} for bound states within a relativistic quantum field theory
by eliminating---according to the spirit of an instantaneous
interaction---any dependence on timelike variables. A central r\^{o}le in
physics is played by the Coulomb potential, a spherically symmetric
potential, i.~e., one which depends only on the radial coordinate $r \equiv
|\vec x|$, parametrized by some coupling strength $\kappa$:
\begin{equation}
V(\vec x) = V_{\rm C}(r) = - \displaystyle\frac{\kappa}{r} \quad , \qquad
\kappa > 0 \quad .
\label{eq:coulpot}
\end{equation}
The bound-state problem defined by the semirelativistic Hamiltonian
(\ref{eq:semrelham}) with the Coulomb potential (\ref{eq:coulpot}) is what we
call {}\kern-.10em\hbox{``}spinless relativistic Coulomb problem."

Over the past years, this spinless relativistic Coulomb problem, in both one-
and two-particle formulation, has been the subject of intense
study.%
\footnote{\normalsize\ The semirelativistic Coulombic Hamiltonians for one-
and two-particle problem, $H^{(1)} = \sqrt{\vec P\,{}^2 + M^2} - \alpha/R$
and $H^{(2)} = 2\sqrt{\vec p\,{}^2 + m^2} - \kappa/r$, respectively, may be
easily equated. Relate their respective phase-space variables $(\vec X,\vec
P)$ and $(\vec x,\vec p\,)$ by re-scaling them by some scale factor $\lambda$
according to $\vec P = \lambda\,\vec p$ and $\vec X = \vec x/\lambda$, which
preserves their fundamental commutation relations: $[\vec X,\vec P] = [\vec
x,\vec p\,]$. Let $\lambda = 2$ and identify both mass and Coulomb coupling
strength parameters according to $M = 2\,m$ and $\alpha = \kappa/2$.} First
of all, Herbst \cite{herbst77}, in a rigorous mathematical discussion,
developed the complete spectral theory of the one-particle counterpart of the
operator (\ref{eq:semrelham}), (\ref{eq:coulpot})---from which one may
directly deduce for the two-particle relativistic
{}\kern-.10em\hbox{``}Coulombic" Hamiltonian under consideration its
essential self-adjointness for $\kappa \le 1$ and the existence of its
Friedrichs extension up to the critical value $\kappa_{\rm cr} = 4/\pi$ of
the coupling constant---and derived a strict lower bound on the energy $E_0$
of the ground state which translates in the two-particle case to
\begin{equation}
E_0 \ge 2\,m\sqrt{1 - \left(\displaystyle\frac{\pi\,\kappa}{4}\right)^2}
\qquad \mbox{for} \quad \kappa < \displaystyle\frac{4}{\pi} \quad .
\label{eq:e0>herbst}
\end{equation}
Durand and Durand \cite{durand83} obtained in a certainly rather involved
analysis of the spinless relativistic Coulomb problem for the particular case
of vanishing orbital angular momentum of the bound-state constituents, as a
by-product of the explicit construction of the corresponding wave function, a
closed analytic expression for the exact energy eigenvalues. Castorina and
co-workers \cite{castorina84} generalized, for small relative distances of
the bound-state constituents, this wave function to arbitrary values of the
orbital angular momentum. Hardekopf and Sucher \cite{hardekopf85} performed a
comprehensive numerical analysis of one- and two-particle relativistic wave
equations for both spin-$0$ as well as spin-$\frac{1}{2}$ particles in the
course of which they were able to show that the result reported in Ref.
\cite{durand83} for the energy eigenvalues must necessarily be wrong. Martin
and Roy \cite{martin89} improved the lower bound (\ref{eq:e0>herbst})
somewhat to
\begin{equation}
E_0 \ge 2\,m\sqrt{\displaystyle\frac{1 + \sqrt{1 - \kappa^2}}{2}}
\qquad \mbox{for} \quad \kappa < 1 \quad .
\label{eq:e0>martin}
\end{equation}
Finally, Raynal and co-workers \cite{raynal94} succeeded in restricting
numerically the ground-state energy eigenvalue of the semirelativistic
Hamiltonian (\ref{eq:semrelham}), (\ref{eq:coulpot}), considered as a
function of the coupling strength $\kappa$, to some remarkably narrow band.

The aim of the present note is to find some simple and, in any case,
analytically given upper bound on the ground-state energy level of the above
semirelativistic {}\kern-.10em\hbox{``}Coulombic" Hamiltonian.

\section{A Variational Upper Bound}

In order to derive an analytic upper bound on the energy eigenvalue of the
ground state of the spinless relativistic Coulomb problem we make use of a
rather standard variational technique. The basic idea of this variational
technique is
\begin{itemize}
\item to calculate the expectation values of the Hamiltonian $H$ under
consideration with respect to a suitably chosen set of trial states
$|\lambda\rangle$ distinguished from each other by some variational parameter
$\lambda$, which yields the $\lambda$-dependent expression $E(\lambda) \equiv
\langle\lambda|H|\lambda\rangle$, and
\item to minimize $E(\lambda)$ with respect to $\lambda$ in order to obtain
the upper bound to the proper energy eigenvalue $E$ of the Hamiltonian $H$ in
the Hilbert-space subsector of the employed trial states $|\lambda\rangle$ as
the above $\lambda$-dependent expression $E(\lambda)$ evaluated at the point
of the minimizing value $\lambda_{\rm min}$ of the variational parameter: $E
\le E(\lambda_{\rm min})$.
\end{itemize}
For the Coulomb potential, the most reasonable choice of trial states is
obviously the one for which the coordinate-space representation $\psi(\vec
x)$ of the states $|\lambda\rangle$ for vanishing radial and orbital angular
momentum quantum numbers is given by the hydrogen-like trial functions
($\lambda > 0$)
$$
\psi(\vec x) = \sqrt{\displaystyle\frac{\lambda^3}{\pi}}\,\exp(- \lambda\,r)
\quad .
$$
For this particular set of trial functions we obtain for the expectation
values we shall be interested in, namely, the ones of the square of the
momentum $\vec p$ and of the inverse of the radial coordinate $r$,
respectively, with respect to the trial states $|\lambda\rangle$
$$
\left\langle\lambda\left|\vec p\,{}^2\right|\lambda\right\rangle = \lambda^2
$$
and
$$
\left\langle\lambda\left|\displaystyle\frac{1}{r}\right|\lambda\right\rangle
= \lambda \quad .
$$

Let us follow this line of arguments in some detail. As an immediate
consequence of the fundamental postulates of any quantum theory, the
expectation value of a given Hamiltonian $H$ taken with respect to any
normalized Hilbert-space state and therefore, in particular, taken with
respect to any of the above trial states must necessarily be larger than or
equal to that eigenvalue $E_0$ of the Hamiltonian $H$ which corresponds to
its ground state:
$$
E_0 \le E(\lambda) \equiv \langle\lambda|H|\lambda\rangle \quad .
$$
Application to the semirelativistic Hamiltonian of Eq. (\ref{eq:semrelham})
yields for the right-hand side of this inequality
$$
E(\lambda) =
2\left\langle\lambda\left|\sqrt{\vec p\,{}^2 + m^2}\right|\lambda\right\rangle
+ \langle\lambda|V(\vec x)|\lambda\rangle \quad .
$$

Here, the rather cumbersome although (for convenient trial states) not
impossible evaluation of the expectation value of the square-root operator
may be very easily circumvented by taking advantage of some trivial but
nevertheless fundamental inequality. This inequality relates the expectation
values, taken with respect to arbitrary Hilbert-space vectors $|\rangle$
normalized to unity, of both the first and second powers of a self-adjoint
but otherwise arbitrary operator ${\cal O} = {\cal O}^\dagger$; it reads
$$
|\langle{\cal O}\rangle| \le \sqrt{\langle{\cal O}^2\rangle} \quad .
$$
For the purposes of the present discussion it is sufficient to replace, in
turn, $E(\lambda)$ by its upper bound obtained by applying this inequality:
$$
E(\lambda) \le
2\sqrt{\left\langle\lambda\left|\vec p\,{}^2\right|\lambda\right\rangle + m^2}
+ \langle\lambda|V(\vec x)|\lambda\rangle \quad .
$$
Identifying in this---as far as its evaluation is concerned,
simplified---upper bound the up to now general potential $V(\vec x)$ with the
Coulomb potential (\ref{eq:coulpot}) and inserting both of the
$\lambda$-dependent expectation values given above implies
\begin{equation}
E(\lambda) \le 2\sqrt{\lambda^2 + m^2} - \kappa\,\lambda \quad .
\label{eq:simpupp}
\end{equation}

{}From this intermediate result, by inspection of the limit $\lambda\to\infty$,
we may state already at this very early stage that, for the semirelativistic
Hamiltonian (\ref{eq:semrelham}), (\ref{eq:coulpot}) to be bounded from below
at all, the Coulombic coupling strength $\kappa$ has to stay below a certain
critical value: $\kappa \le 2$.

The value of the variational parameter $\lambda$ which minimizes the upper
bound on the right-hand side of Eq. (\ref{eq:simpupp}) may be determined from
the derivative of this expression with respect to $\lambda$:
$$
\lambda_{\rm min}
= \displaystyle\frac{m\,\kappa}{2\sqrt{1 - \displaystyle\frac{\kappa^2}{4}}}
\quad .
$$
For this value of $\lambda$, by shuffling together all our previous
inequalities, we find that the energy eigenvalue corresponding to the ground
state of the semirelativistic Hamiltonian (\ref{eq:semrelham}) with Coulomb
potential (\ref{eq:coulpot}), $E_0$, is bounded from above by
\begin{equation}
E_0 \le 2\,m\sqrt{1 - \displaystyle\frac{\kappa^2}{4}} \quad .
\label{eq:groundupp}
\end{equation}
The reality of this upper bound requires again $\kappa \le 2$.

\section{Summary}

Within the framework of a simple variational technique we derived an analytic
upper bound, Eq. (\ref{eq:groundupp}), to the ground-state energy level of
the spinless relativistic Coulomb problem. However, for any nonvanishing
value of the Coulombic coupling strength $\kappa$, this upper bound
(\ref{eq:groundupp}) turns out, in fact, to be violated by the lowest energy
eigenvalue obtainable from the analytic expression given in Ref.
\cite{durand83}. This provides, of course, further confirmation of the
corresponding findings of Ref. \cite{hardekopf85}; a similar observation has
been made in Ref. \cite{leyaouanc94}.


\newpage

\normalsize
\begin{thebibliography}{99}
\bibitem{salpeter52} {\sc E. E. Salpeter}, Phys. Rev. {\bf 87} (1952) 328.
\bibitem{salpeter51} {\sc E. E. Salpeter} and {\sc H. A. Bethe}, Phys. Rev.
{\bf 84} (1951) 1232.
\bibitem{herbst77} {\sc I. W. Herbst}, Commun. Math. Phys. {\bf 53} (1977)
285; {\bf 55} (1977) 316 (addendum).
\bibitem{durand83} {\sc B. Durand} and {\sc L. Durand}, Phys. Rev. D {\bf 28}
(1983) 396; {\bf ??} (1994) ???? (erratum).
\bibitem{castorina84} {\sc P. Castorina}, {\sc P. Cea}, {\sc G. Nardulli},
and {\sc G. Paiano}, Phys. Rev. D {\bf 29} (1984) 2660.
\bibitem{hardekopf85} {\sc G. Hardekopf} and {\sc J. Sucher}, Phys. Rev. A
{\bf 31} (1985) 2020.
\bibitem{martin89} {\sc A. Martin} and {\sc S. M. Roy}, Phys. Lett. B {\bf
233} (1989) 407.
\bibitem{raynal94} {\sc J. C. Raynal}, {\sc S. M. Roy}, {\sc V. Singh}, {\sc
A. Martin}, and {\sc J. Stubbe}, Phys. Lett. B {\bf 320} (1994) 105.
\bibitem{leyaouanc94} {\sc A. Le Yaouanc}, {\sc L. Oliver}, and {\sc J.-C.
Raynal}, Orsay preprint LPTHE Orsay 93/43 (1993).
\end{thebibliography}
\end{document}